\documentstyle[12pt]{article}
\begin{document}
\tolerance=5000
\def\pp{{\, \mid \hskip -1.5mm =}}
\def\cL{{\cal L}}
\def\be{\begin{equation}}
\def\ee{\end{equation}}
\def\bea{\begin{eqnarray}}
\def\eea{\end{eqnarray}}
\def\tr{{\rm tr}\, }
\def\nn{\nonumber \\}
\def\e{{\rm e}}
\def\D{{D \hskip -3mm /\,}}

\def\SEH{S_{\rm EH}}
\def\SGH{S_{\rm GH}}
\def\AdS5{{{\rm AdS}_5}}
\def\S4{{{\rm S}_4}}
\def\gfv{{g_{(5)}}}
\def\gfr{{g_{(4)}}}
\def\SC{{S_{\rm C}}}
\def\RH{{R_{\rm H}}}

\def\wlBox{\mbox{
\raisebox{0.1cm}{$\widetilde{\mbox{\raisebox{-0.1cm}\fbox{\ }}}$}}}
\def\htBox{\mbox{
\raisebox{0.1cm}{$\hat{\mbox{\raisebox{-0.1cm}{$\Box$}}}$}}}

\  \hfill
\begin{minipage}{3.5cm}
March 2001 \\
\end{minipage}

\vfill

\begin{center}
{\large\bf AdS/CFT and quantum-corrected brane entropy}

\vfill

{\sc Shin'ichi NOJIRI}\footnote{nojiri@cc.nda.ac.jp},
and {\sc Sergei D. ODINTSOV}$^{\spadesuit}$\footnote{
odintsov@ifug5.ugto.mx, odintsov@mail.tomsknet.ru}\\

\vfill

{\sl Department of Applied Physics \\
National Defence Academy,
Hashirimizu Yokosuka 239-8686, JAPAN}

\vfill

{\sl $\spadesuit$
Instituto de Fisica de la Universidad de Guanajuato,
Lomas del Bosque 103, Apdo. Postal E-143, 
37150 Leon,Gto., MEXICO 
and Tomsk State Pedagogical University, 634041 Tomsk, RUSSIA}

\vfill

{\bf ABSTRACT}

\end{center}

It is shown that quantum-induced (inflationary) brane Universe occurs 
in the bulk 5d AdS black hole in accordance with AdS/CFT correspondence.
Brane stress tensor is induced by quantum effects of dual CFT and brane
crosses
the horizon of AdS black hole. Quantum-corrected Hubble constant,
Hawking temperature and entropy are found on the brane (and at the horizon).
The similarity between CFT entropy at the horizon and FRW equations is
extended
on the quantum level. This suggests the way to understand cosmological 
entropy bounds in quantum gravity.

\newpage

\section{Introduction}

It is quite well-known fact that holographic principle suggests 
the interesting bounds between microscopic and Bekenstein-Hawking 
entropy \cite{H} as it was discussed in refs.\cite{HS,HMS}.
Recently, the very interesting attempt to study the holographic 
principle in Friedmann-Robertson-Walker (FRW) Universe 
filled by 
conformal matter 
has been done by Verlinde \cite{EV}.
Using dual AdS-description \cite{AdS} it has been found the relation 
between entropy (energy) of CFT and cosmological equations controlling 
the behaviour of scale factor in FRW Universe.
In particular, the equation controlling the entropy bounds 
during evolution has been obtained \cite{EV} and Cardy-Verlinde 
formula has been derived.
These results have been subsequently generalized and discussed in 
a number of works \cite{related,NO}.

From another side there is some interest to the cosmological brane 
universe realized as some kind of the boundary in the AdS-Schwarzschild 
black hole as it was discussed in \cite{Gubser}. Related with 
the above works about the holographic principle in FRW 
universe, one interesting 
extension has been presented in ref.\cite{SV} where similar 
questions about the cosmological entropy, evolution, etc. within the 
holographic principle have been studied from classical brane-world 
perspective\cite{RS}. In particular, the behaviour of
the CFT entropy at the horizon of bulk 5d AdS BH has been 
investigated and its comparison with FRW equations has been done.

In the study of brane-worlds and their applications two main approaches 
could be considered. In the first, more traditional approach one starts 
from the
higher dimensional theory which gives the higher dimensional bulk solution
(say, AdS space). The next step is to get the necessary brane universe.
In order to achieve this one adds by hands some boundary terms (brane 
vacuum energy). In this way, almost any brane universe may be easily 
obtained. 

There is, however, another way which is closely connected with AdS/CFT 
correspondence and quantum properties of the system under discussion.
In this, second approach the bulk action is not modified. However, 
the boundary terms are not fine-tuned, they are predicted by some 
reasonable assumptions. First of all, part of surface terms represents 
the Gibbons-Hawking term which is responsible for getting the variational 
procedure to be well-defined.
Second contribution to surface terms comes from the principle that 
leading divergence of bulk space (say, of AdS) should be cancelled. 
Final part is 
the dynamical one: it is produced by quantum effects of CFT on the brane.
After having such action the brane universe comes as the solution of 
equations of motion. Definitely, very few brane universes naturally appear
as a result of such dynamical solution of equations of motion.
In this way, so-called Brane New World \cite{HHR,NObr} has been constructed.

 The purpose of the present paper is to 
generalize the situation 
described in ref.\cite{SV} to the case of above quantum-induced (or AdS/CFT 
induced) brane-worlds suggested in refs.\cite{HHR,NObr}, 
where the quantum creation of the brane universe is discussed. 
In this way, from one side one gets quantum-corrected FRW Universe 
equations as they look from the point of view of not only 
brane observer  but also 
from the point of view of quantum induced brane-world.
 From another side, one gets the quantum-corrected brane entropy 
as well as Hubble constant and Hawking temperature 
at the horizon. Finally, this may be considered as 
extension of scenario of refs.\cite{HHR,NObr} 
(see refs.\cite{extension} for related questions) 
which admits also generalization for the presence 
of non-trivial dilaton and (or) supersymmetrization \cite{NOO} 
for the case when brane crosses the horizon of AdS-black hole.

\section{Brane New World in AdS-Schwarzschild Black Hole}

We assume the brane connects two bulk spaces and we may also 
identify the two bulk spaces as in \cite{RS} by imposing $Z_2$ 
symmetry. 
We start with the Minkowski signature action $S$ which is 
the sum of the Einstein-Hilbert action $\SEH$ with the 
cosmological term, the Gibbons-Hawking surface term $\SGH$, 
the surface counter term $S_1$ and the trace anomaly induced action 
${\cal W}$: 
\bea
\label{Stotal}
S&=&\SEH + \SGH + 2 S_1 + {\cal W}, \\
\label{SEHi}
\SEH&=&{1 \over 16\pi G}\int d^5 x \sqrt{-\gfv}\left(R_{(5)} 
 + {12 \over l^2} \right), \\
\label{GHi}
\SGH&=&{1 \over 8\pi G}\int d^4 x \sqrt{-\gfr}\nabla_\mu n^\mu, \\
\label{S1}
S_1&=& -{6 \over 16\pi G l}\int d^4 x \sqrt{\gfr} , \\
\label{W}
{\cal W}&=& b \int d^4x \sqrt{-\widetilde g}\widetilde F A 
 + b' \int d^4x\sqrt{\widetilde g}
\left\{A \left[2{\wlBox}^2 
+\widetilde R_{\mu\nu}\widetilde\nabla_\mu\widetilde\nabla_\nu 
\right.\right. \nn
&& \left.\left. - {4 \over 3}\widetilde R \wlBox^2 
+ {2 \over 3}(\widetilde\nabla^\mu \widetilde R)\widetilde\nabla_\mu
\right]A + \left(\widetilde G - {2 \over 3}\wlBox \widetilde R
\right)A \right\} \\
&& -{1 \over 12}\left\{b''+ {2 \over 3}(b + b')\right\}
\int d^4x \sqrt{\widetilde g} 
\left[ \widetilde R - 6\wlBox A 
 - 6 (\widetilde\nabla_\mu A)(\widetilde \nabla^\mu A)
\right]^2 \nn
 .\nonumber
\eea 
Here the quantities in the  5 dimensional bulk spacetime are 
specified by the suffices $_{(5)}$ and those in the boundary 4 
dimensional spacetime are specified by $_{(4)}$ (for details, 
see \cite{NObr}). 
In (\ref{GHi}), $n^\mu$ is the unit vector normal to the 
boundary. The Gibbons-Hawking term $S_{\rm GH}$ is necessary 
in order to make the variational method well-defined when there is 
boundary in the spacetime. In (\ref{S1}), the coefficient of 
$S_1$ is determined from AdS/CFT 
\cite{HHR}. The factor 2 in front of $S_1$ 
is coming from that we have two bulk regions which 
are connected with each other by the brane. 
In (\ref{W}), one chooses the 4 dimensional boundary metric as 
$\gfr_{\mu\nu}=\e^{2A}\tilde g_{\mu\nu}$, where 
$\tilde g_{\mu\nu}$ is a reference metric.  
$G$ ($\tilde G$) and $F$ ($\tilde F$) are the 
Gauss-Bonnet invariant and the square of the Weyl tensor. 
${\cal W}$ can be obtained by integrating the conformal 
anomaly with respect to the scale factor $A$ of the metric tensor 
since the conformal anomaly should be given by the variation of 
the quantum effective action with respect to $A$.
Note that quantum effects of brane CFT are taken into account via 
Eq.(\ref{W})\footnote{In \cite{NOZcas}, the bulk gravitational 
Casimir effect has been considered and it has been found that the 
Casimir effect leads to deformation of 5d AdS space shape as well 
as of shape of branes. The account of bulk quantum effects, 
however, do not change the qualitative picture and the brane 
inflation still occurs. Then the role of bulk scalar quantum 
effect is not relevant in present context.
}.

In the effective action (\ref{W}) induced by brane quantum conformal 
matter, in general, with $N$ scalar, $N_{1/2}$ spinor, $N_1$ vector 
fields, $N_2$  ($=0$ or $1$) gravitons and $N_{\rm HD}$ higher 
derivative conformal scalars, $b$, $b'$ and $b''$ are \cite{NObr}
\bea
\label{bs}
&& b={N +6N_{1/2}+12N_1 + 611 N_2 - 8N_{\rm HD} 
\over 120(4\pi)^2}\nn 
&& b'=-{N+11N_{1/2}+62N_1 + 1411 N_2 -28 N_{\rm HD} 
\over 360(4\pi)^2}\ , \quad b''=0\ .
\eea
For typical examples motivated by AdS/CFT correspondence 
one has:

\noindent
a) ${\cal N}=4$ $SU(N)$ SYM theory 
\be
\label{N4bb}
b=-b'={N^2 -1 \over 4(4\pi )^2}\ ,
\ee 
b) ${\cal N}=2$ $Sp(N)$ theory 
\be
\label{N2bb}
b={12 N^2 + 18 N -2 \over 24(4\pi)^2}\ ,\quad 
b'=-{12 N^2 + 12 N -1 \over 24(4\pi)^2}\ .
\ee
 Note that $b'$ is negative in the above cases. It is important to note
that brane quantum gravity may be taken into account via the
contribution to correspondent parameters $b,b'$.

Then on the brane, we have the following equation which generalizes the 
classical brane equation of the motion:
\bea
\label{eq2b}
0&=&{48 l^4 \over 16\pi G}\left(A_{,z} 
 - {1 \over l}\right)\e^{4A}
+b'\left(4 \partial_\tau^4 A + 16 \partial_\tau^2 A\right) \nn
&& - 4(b+b')\left(\partial_\tau^4 A - 2 \partial_\tau^2 A 
 - 6 (\partial_\tau A)^2\partial_\tau^2 A \right) \ .
\eea
This equation is derived from the condition that the variation 
of the action on the brane, or the boundary of the bulk spacetime, 
vanishes under the variation over $A$. The first term 
proportional to $A_{,z}$ expresses the bulk gravity force acting 
on the brane and the term proportional to ${1 \over l}$ comes 
from the brane tension. The terms containing $b$ or $b'$ 
express the contribution from the conformal anomaly induced effective action 
(quantum effects).
In (\ref{eq2b}), one uses the form of the metric as 
\be
\label{metric1}
ds^2=dz^2 + \e^{2A(z,\tau)}\tilde g_{\mu\nu}dx^\mu dx^\nu\ ,
\quad \tilde g_{\mu\nu}dx^\mu dx^\nu\equiv l^2\left(-d \tau^2 
+ d\Omega^2_3\right)\ .
\ee
Here $d\Omega^2_3$ corresponds to the metric of 3 dimensional 
unit sphere. 

As a bulk space, we consider 5d AdS-Schwarzschild black hole spacetime, 
whose metric is given by,
\be
\label{AdSS}
ds_{\rm AdS-S}^2 = {1 \over h(a)}da^2 - h(a)dt^2 
+ a^2 d\Omega_3^2 \ ,\ \ 
h(a)= {a^2 \over l^2} + 1 - {16\pi GM \over 3 V_3 a^2}\ .
\ee
Here $V_3$ is the volume of the unit 3 sphere.  
If one chooses new coordinates $(z,\tau)$ by
\bea
\label{cc1}
&& {\e^{2A} \over h(a)}A_{,z}^2 - h(a) t_{,z}^2 = 1 \ ,
\quad {\e^{2A} \over h(a)}A_{,z}A_{,\tau} - h(a)t_{,z} t_{,\tau}
= 0 \nn
&& {\e^{2A} \over h(a)}A_{,\tau}^2 - h(a) t_{,\tau}^2 
= -\e^{2A}\ .
\eea
the metric takes the warped form (\ref{metric1}). Here $a=l\e^A$.
In general we might be unable to  rewrite globally the metric in 
(\ref{AdSS}) in the form of (\ref{metric1}). Nevertheless, it can be done 
in the neighbourhood of the brane, what is necessary here. 
Further choosing a coordinate $\tilde t$ by 
$d\tilde t = l\e^A d\tau$, the metric on the brane takes FRW form: 
\be
\label{e3}
\e^{2A}\tilde g_{\mu\nu}dx^\mu dx^\nu= -d \tilde t^2  
+ l^2\e^{2A} d\Omega^2_3\ .
\ee
By solving Eqs.(\ref{cc1}), we have
\be
\label{e4}
H^2 = A_{,z}^2 - h\e^{-2A}= A_{,z}^2 - {1 \over l^2}
 - {1 \over a^2} + {16\pi GM \over 3 V_3 a^4}\ .
\ee
Here the Hubble constant $H$ is introduced:
$H={dA \over d\tilde t}$. 
On the other hand, from (\ref{eq2b}) one gets
\bea
\label{e6}
A_{,z}&=&{1 \over l} + {\pi G \over 3}\left\{ 
-4b'\left(\left(H_{\tilde t \tilde t \tilde t} + 4 H_{\tilde t}^2 
+ 7 H H_{\tilde t\tilde t} + 18 H^2 H_{\tilde t} + 6H^4\right) 
\right.\right. \nn
&& \left. + {4 \over a^2} \left(H_{\tilde t} 
+ H^2\right)\right) 
+ 4(b+b') \left(\left(H_{\tilde t \tilde t \tilde t} 
+ 4 H_{\tilde t}^2 \right.\right. \nn
&& \left.\left.\left. + 7 H H_{\tilde t\tilde t} 
+ 12 H^2 H_{\tilde t} \right) 
 - {2 \over a^2} \left(H_{\tilde t} + H^2\right)\right) \right\}\ .
\eea
Then combining (\ref{e4}) and (\ref{e6}), we find
\bea
\label{e7}
&& H^2 = - {1 \over l^2}
 - {1 \over a^2} + {16\pi GM \over 3 V_3 a^4} 
+ \left[{1 \over l} + {\pi G \over 3}\left\{ 
-4b'\left(\left(H_{,\tilde t \tilde t \tilde t} + 4 H_{,\tilde t}^2 
\right.\right.\right.\right. \nn
&& \ \left.\left. + 7 H H_{,\tilde t\tilde t} 
+ 18 H^2 H_{,\tilde t} + 6 H^4\right) 
+ {4 \over a^2} \left(H_{,\tilde t} + H^2\right)\right) \\
&& \ + 4(b+b') \left(\left(H_{,\tilde t \tilde t \tilde t} 
+ 4 H_{,\tilde t}^2 + 7 H H_{,\tilde t\tilde t} 
+ 12 H^2 H_{,\tilde t} \right) 
\left.\left. - {2 \over a^2} \left(H_{,\tilde t} 
+ H^2\right)\right) \right\}\right]^2\ .\nonumber
\eea
This expresses the quantum correction to the corresponding brane  
equation in \cite{SV}. In fact, if we put $b=b'=0$, 
Eq.(\ref{e7}) reduces to the classical FRW equation
\be
\label{e8}
H^2 = - {1 \over a^2} + {16\pi GM \over 3 V_3 a^4} \ .
\ee 
Further by differentiating Eq.(\ref{e7}) with respect to 
$\tilde t$, we obtain
\bea
\label{e9}
&& H_{,\tilde t} =  {1 \over a^2} - {32\pi GM \over 3 V_3 a^4} 
+ {\pi G \over 3 H}\left[{1 \over l} 
+ {\pi G \over 3}\left\{ 
-4b'\left(\left(H_{,\tilde t \tilde t \tilde t} + 4 H_{,\tilde t}^2 
\right.\right.\right.\right. \nn
&& \ \left. \left. + 7 H H_{,\tilde t\tilde t} 
+ 18 H^2 H_{,\tilde t} + 6 H^4\right) 
+ {4 \over a^2} \left(H_{,\tilde t} + H^2
\right)\right) \nn
&& \ + 4(b+b') \left(\left(H_{,\tilde t \tilde t \tilde t} 
+ 4 H_{,\tilde t}^2 + 7 H H_{,\tilde t\tilde t} 
+ 12 H^2 H_{,\tilde t} \right) 
\left.\left. - {2 \over a^2} \left(H_{,\tilde t} 
+ H^2\right) \right)\right\}\right] \nn
&& \ \times \left\{ 
-4b'\left(\left(H_{,\tilde t \tilde t \tilde t \tilde t} 
+ 15 H_{,\tilde t} H_{\tilde t\tilde t} 
+ 7 H H_{,\tilde t\tilde t\tilde t} + 18 H^2 H_{,\tilde t\tilde t} 
\right.\right.\right. \nn
&& \ \left.\left. + 36 H H_{,\tilde t}^2 
+ 24 H^3 H_{,\tilde t} \right) + {4 \over a^2} 
\left(H_{,\tilde t\tilde t}  - 2 H^3\right) \right) \nn
&& \ + 4(b+b') \left(\left(H_{,\tilde t \tilde t \tilde t \tilde t} 
+ 15 H_{,\tilde t} H_{,\tilde t\tilde t} 
+ 7 H H_{,\tilde t\tilde t\tilde t} 
+ 12 H^2 H_{,\tilde t\tilde t}\right.\right. \nn
&& \ \left.\left.\left. + 24 H H_{,\tilde t}^2 \right) 
 - {2 \over a^2} \left(H_{,\tilde t\tilde t} 
 - 2H^2\right)\right) \right\}\ .
\eea
One can rewrite the above equations (\ref{e7}) and (\ref{e9}) 
in the form of  FRW equations: 
\bea
\label{e10}
&& H^2 = - {1 \over a^2} 
+ {8\pi G_4 \rho \over 3} \\
\label{e10b}
&& \rho={l \over a}\left[ {M \over V_3 a^3} \right. 
+ {3a \over 16\pi G}\left[
\left[{1 \over l} + {\pi G \over 3}\left\{ 
-4b'\left(\left(H_{,\tilde t \tilde t \tilde t} + 4 H_{,\tilde t}^2 
+ 7 H H_{,\tilde t\tilde t} \right.\right.\right.\right. \right.\nn
&& \ \left.\left. + 18 H^2 H_{,\tilde t} + 6 H^4\right) 
+ {4 \over a^2} \left(H_{,\tilde t} + H^2\right)\right) 
+ 4(b+b') \left(\left(H_{,\tilde t \tilde t \tilde t} 
+ 4 H_{,\tilde t}^2 \right. \right. \nn
&& \ \left.\left.\left.\left.\left.\left. + 7 H H_{,\tilde t\tilde t} 
+ 12 H^2 H_{,\tilde t} \right) - {2 \over a^2} 
\left(H_{,\tilde t} + H^2\right)\right) \right\}\right]^2
 - {1 \over l^2} \right]\right]\ ,\\
\label{e11}
&& H_{,\tilde t} =  {1 \over a^2} - 4\pi G_4(\rho + p) \\
\label{e11b}
&& \rho + p= {l \over a}\left[
 {4 M \over 3 V_3 a^3} \right. - {1 \over 24l^3 H}\left[{1 \over l} 
+ {\pi G \over 3}\left\{ 
-4b'\left(\left(H_{,\tilde t \tilde t \tilde t} + 4 H_{,\tilde t}^2 
+ 7 H H_{,\tilde t\tilde t} \right.\right.\right.\right. \nn
&& \ \left.\left. + 18 H^2 H_{,\tilde t} + 6 H^4\right) 
+ {4 \over a^2} \left(H_{,\tilde t} + H^2
\right)\right) \nn
&& \ + 4(b+b') \left(\left(H_{,\tilde t \tilde t \tilde t} 
+ 4 H_{,\tilde t}^2 + 7 H H_{,\tilde t\tilde t} 
+ 12 H^2 H_{,\tilde t} \right) 
\left.\left. - {2 \over a^2} \left(H_{,\tilde t} + H^2\right) 
\right)\right\}\right] \nn
&& \ \times \left\{ 
-4b'\left(\left(H_{,\tilde t \tilde t \tilde t \tilde t} 
+ 15 H_{,\tilde t} H_{\tilde t\tilde t} 
+ 7 H H_{,\tilde t\tilde t\tilde t} 
+ 18 H^2 H_{,\tilde t\tilde t} 
+ 36 H H_{,\tilde t}^2 \right.\right.\right. \nn
&& \ \left.\left. + 24 H^3 H_{,\tilde t} \right)
+ {4 \over a^2} \left(H_{,\tilde t\tilde t} - 2 H^3\right) \right) 
+ 4(b+b') \left(\left(H_{,\tilde t \tilde t \tilde t \tilde t} 
+ 15 H_{,\tilde t} H_{,\tilde t\tilde t} \right.\right. \nn
&& \ \left.\left. 
+ 7 H H_{,\tilde t\tilde t\tilde t} + 12 H^2 H_{,\tilde t\tilde t}
+ 24 H H_{,\tilde t}^2 \right)
\left.\left. - {2 \over a^2} \left(H_{,\tilde t\tilde t} 
 - 2H^2\right)\right) \right\}\right]\ .
\eea
Here 4d Newton constant $G_4$ is given by
\be
\label{e12}
G_4={2G \over l}\ .
\ee
and quantum corrections from CFT are included into the definition of
energy (pressure). These quantum corrected FRW equations are written
from quantum-induced brane-world perspective. Similar equations from the
point 
of view of 4d brane observer (who does not know about 5d AdS bulk)
have been presented in ref.\cite{NO}.
Clearly, brane-world approach gives more information.
As the correction terms include higher derivatives, these 
terms become relevant when the universe changes its size very 
rapidly as in the very early universe. 

It is not so clear if the energy density $\rho$ and the 
pressure $p$ satisfy the energy conditions  from the expressions 
in (\ref{e10b}) and (\ref{e11b}), because quantum effects generally
may violate the energy conditions. For the solution of (\ref{e10}), 
however, $\rho$ is always positive since (\ref{e10}) can be 
rewritten
\be
\label{e10bb}
\rho= {3 \over 8\pi G_4}\left(H^2 + {1 \over a^2}\right) >0. 
\ee
We also have from (\ref{e11})
\be
\label{e11bb}
\rho+p={1 \over 4\pi G_4}\left({1 \over a^2} - H_{,\tilde t}
\right)\ .
\ee
Therefore the weak energy condition should be 
satisfied if ${1 \over a^2} - H_{,\tilde t}>0$ in the solution. 
In order to clarify the situation, we consider the specific case 
of $b+b'=0$ as in ${\cal N}=4$ theory and we assume that $b'$ is 
small. Then from (\ref{e10}) and (\ref{e11}) and by 
differentiating (\ref{e11}) with respect $\tilde t$, one gets
\bea
\label{aa1}
&& H^2=-{1 \over a^2}+{8\pi G_4 Ml \over 3 V_3 a^4}
+{\cal O}\left(b'\right)\ ,\quad
H_{,\tilde t}={1 \over a^2}-{16\pi G_4 Ml \over 3 V_3 a^4}
+{\cal O}\left(b'\right)\ ,\nn
&& H_{,\tilde t\tilde t}=-{2 \over a^2}H
+{64\pi G_4 Ml \over 3 V_3 a^4}H
+{\cal O}\left(b'\right)\ ,\quad \mbox{etc.}
\eea
Then by using (\ref{e10b}) and (\ref{e11b}), we find
\bea
\label{aa2}
\rho&=&{Ml \over V_3 a^4}-{b' \over 2}\left(
{8\pi G_4 Ml \over V_3 a^6}
 - {128\pi^2 G_4^2 M^2l^2 \over 3 V_3^2 a^8}\right)
+{\cal O}\left({b'}^2\right)\ ,\nn
p&=&{Ml \over 3V_3 a^4}-{b' \over 2}\left(
{8\pi G_4 Ml \over V_3 a^6}
 - {640\pi^2 G_4^2 M^2l^2 \over 9 V_3^2 a^8}\right)
+{\cal O}\left({b'}^2\right)\ .
\eea
The correction part of $\rho$ is not always positive but $\rho$ 
itself should be positive, what is clear from (\ref{e10bb}). 
One also gets
\be
\label{aa3}
\rho+p={4Ml \over 3V_3 a^4}-{b' \over 2}\left(
{16\pi G_4 Ml \over V_3 a^6}
 - {1024\pi^2 G_4^2 M^2l^2 \over 9 V_3^2 a^8}\right)
+{\cal O}\left({b'}^2\right)\ .
\ee 
Then the correction part seems to be not always positive and 
the weak energy condition might be broken. 
As the above discussion is based on the perturbation theory, we will 
discuss the weak energy condition later  using the de Sitter 
type brane universe solution. 

Let us consider the solution of quantum-corrected FRW equation 
(\ref{e7}). Assume the de Sitter type solution 
\be
\label{dS1}
a=A\cosh B\tilde t\ .
\ee
Substituting (\ref{dS1}) into (\ref{e7}), 
one finds the following equations should be satisfied:
\bea
\label{dS3} 
0&=& - {1 \over B^2} - {1 \over l^2} 
+ \left({1 \over l} - 8\pi G b' B^4 \right)^2 \\
\label{dS4}
0&=& B^2 - {1 \over A^2} 
+ 2 \left({1 \over l} - 8\pi G b' B^4 \right)
{\pi G \over 3} \left(24b' + 8b\right) 
\left(B^4 - {B^2 \over A^2}\right) \\
\label{dS5}
0&=& {16\pi GM \over 3V_3} 
+ \left({\pi G \over 3}\right)^2 \left(24b' + 8b\right)^2 
\left(B^4 - {B^2 \over A^2}\right)^2 \ .
\eea
Eq.(\ref{dS3}) tells that there is no de Sitter type solution 
if there is no quantum correction, or if $b'=0$.  
Eq.(\ref{dS5}) tells that if the black hole mass $M$ is 
non-vanishing and positive, there is no any solution 
of the de Sitter-like brane. When $M=0$, Eqs.(\ref{dS4}) 
and (\ref{dS5}) are trivially satisfied if 
$A^2={1 \over B^2}$. Actually this case corresponds to  well-known 
anomaly-driven 
inflation \cite{S} (for recent discussion, see \cite{HHR2}). 
Eq.(\ref{dS3}) has 
unique non-trivial solution for $B^2$, which corresponds 
to the de Sitter brane universe  in \cite{HHR,NObr}. 

When $M<0$, there is no horizon and the curvature singularity 
becomes naked. We will, however, formally consider the case since 
there is no de Sitter-like brane solution in the classical case 
($b'=0$) even if $M$ is negative. 
If $M\neq 0$ or $A^2\neq {1 \over B^2}$, Eq.(\ref{dS4}) has the 
following form:
\be
\label{dS6}
0= 1 + 2 \left({1 \over l} - {8\pi G b' \over l^4}B^4 \right)
{\pi G \over 3l^4} \left(24b' + 8b\right) B^2 \ .
\ee
Eq.(\ref{dS6}) is not always compatible with Eq.(\ref{dS3}) and 
gives a non-trivial constraint on $G$, $l$, $b$ and $b'$. 
If the constraint is satisfied, $B^2$ can be uniquely 
determined by (\ref{dS3}) or (\ref{dS6}). Then (\ref{dS5}) 
can be solved with respect to $A^2$. 

Now we consider the above constraint and solution for $B^2$. 
By combining (\ref{dS3}) and (\ref{dS6}),  one obtains 
\bea
\label{dS7} 
0&=&B^6 + {1 \over l^2}B^4 - {1 \over \eta}\ ,
\quad \eta\equiv 4\left(24b' + 8b\right)\left({\pi G \over 3}
\right)^2 \\
\label{dS8}
0&=& \left({1 \over l^2} + B^2\right)^3 
 - \left\{{1 \over l} \left({1 \over l^2} + B^2\right)
 - \zeta\right\}\ ,
\quad \zeta \equiv {6b' \over 24b' + 8b}
\left({3 \over \pi G}\right)\ .
\eea
In most of cases, $\eta$ is negative and $\zeta$ is positive. 
The explicit solution of (\ref{dS7}) is given by 
\be
\label{dS9}
B^2=-{1 \over 3l^2} 
+ \left({1 \over 27l^6} - {1 \over 2\eta} + 
\sqrt{{1 \over 4\eta^2} - {1 \over 27l^6\eta}}
\right)^{1 \over 3}
+ \left({1 \over 27l^6} - {1 \over 2\eta} - 
\sqrt{{1 \over 4\eta^2} - {1 \over 27l^6\eta}}
\right)^{1 \over 3} \ .
\ee
On the other hand, if ${\zeta^4 \over 4}
 - {\zeta^3 \over 27l^4}>0$, the solution of (\ref{dS8}) is 
given by
\bea
\label{dS10}
B^2 &=& - {2 \over 3l^2} 
+ \left(-{1 \over 27l^6} + {\zeta \over 3l^3} - {\zeta^2 \over 2} 
+ \sqrt{{\zeta^4 \over 4} - {\zeta^3 \over 27l^4}}
\right)^{1 \over 3} \nn
&& + \left(-{1 \over 27l^6} + {\zeta \over 3l^3} - {\zeta^2 \over 2} 
 - \sqrt{{\zeta^4 \over 4} - {\zeta^3 \over 27l^4}}
\right)^{1 \over 3}\ .
\eea
or if ${\zeta^4 \over 4} - {\zeta^3 \over 27l^4}<0$, 
the solutions are
\be
\label{dS11}
B^2+ {2 \over 3l^2} = \xi + \xi^*\ ,\quad 
\xi\omega + \xi^*\omega^2\ , \quad 
\xi\omega^2 + \xi^*\omega\ .
\ee
Here 
\be
\label{dS12}
\xi = \left(-{1 \over 27l^6} + {\zeta \over 3l^3} - {\zeta^2 \over 2} 
+ i\sqrt{{\zeta^3 \over 27l^4}-{\zeta^4 \over 4}} 
\right)^{1 \over 3}\ ,\quad 
\omega=\e^{2i\pi \over 3}\ .
\ee
Then if the solution (\ref{dS9}) coincides with 
any of the solutions (\ref{dS10}) or (\ref{dS11}), 
there occurs quantum-induced de Sitter-like brane realized in d5 AdS BH. 
In a sense, we got the extension of scenario of refs.\cite{HHR,NObr}
 for quantum-induced brane-worlds within AdS/CFT set-up when
bulk is given by d5 AdS BH.

For the de Sitter type solution (\ref{dS1}), Eq.(\ref{e11bb}) has 
the following form:
\be
\label{e11bbb}
\rho+p={1 \over 4\pi G_4}\left({1 \over A^2} - B^2\right)
{1 \over \cosh^2 B\tilde t}\ .
\ee
Then the weak energy condition can be satisfied if 
\be
\label{e11bbb2}
{1 \over A^2} \geq B^2\ .
\ee
For the exact de Sitter solution corresponding to $M=0$, we 
have ${1 \over A^2} = B^2$ and Eq.(\ref{e11bbb2}) is satisfied. 
For more general solution in (\ref{dS9}) or (\ref{dS11}), 
$B$ and $A$ non-trivially depend on the parameters $G$, $M$, 
$b$ and $b'$ and it is not so clear if Eq.(\ref{e11bbb2}) is 
always satisfied. 

Since the solution whose form is $a=A\cosh B\tilde t$ exists when 
the parameters satisfy a special constraint
and the black hole mass $M$ is negative, 
we now consider 
a perturbation from the de Sitter brane solution by assuming 
that the black hole mass $M$ is small. By choosing $a$ as 
\be
\label{dS13}
\ln a = \ln \left({1 \over B}\cosh B\tilde t\right) + h(\tilde t)\ ,
\ee
we expand (\ref{e7}) in the first order of $M$ and $h$:
\bea
\label{dS14}
0&=&-2B\tanh B\tilde t h_{,\tilde t} + {2B^2 \cosh^2 B\tilde t}
+ {16 \pi GM B^4 \over 3V_3 \cosh^4 B\tilde t} \nn
&& + \left.\left.{2\pi G \over 3}\left({1 \over l} 
- 8\pi G B^4 b'\right)\right\{-4b'\right(
h_{,\tilde t\tilde t\tilde t\tilde t} 
+ 7B\tanh B\tilde t \, h_{,\tilde t\tilde t\tilde t} \nn
&& + B^2\left(18 - {6 \over \cosh^2 B\tilde t}\right)
h_{,\tilde t\tilde t} 
+ B^3\left({24 \sinh B\tilde t \over \cosh B\tilde t} 
+ {12\sinh B\tilde t \over \cosh^3 B\tilde t} \right)
h_{,\tilde t} \nn 
&& \left.\left. - {8B^4 \over \cosh^2 B\tilde t} h\right) 
+ 4(b+b')\right(h_{,\tilde t\tilde t\tilde t\tilde t} 
+ 7B\tanh B\tilde t \,h_{,\tilde t\tilde t\tilde t} \nn
&& \left.\left. + B^2\left(12 - {2 \over \cosh^2 B\tilde t}\right)
h_{,\tilde t\tilde t} + {14 B^3 \sinh B\tilde t \over 
\cosh^3 B\tilde t} h_{,\tilde t}
 - {4B^4 \over \cosh^2 B\tilde t} h\right)\right\}\ .
\eea
When $\tilde t$ is small, the solution of (\ref{dS14}) can 
be given by the power series of $\tilde t$ as 
$h(\tilde t)=\sum_{n=0}^\infty h_n \tilde t^n$ with 
a proper boundary condition. On the other hand,  
when $\tilde t$ is large, Eq.(\ref{dS14}) has the following 
form:
\bea
\label{dS15}
0&=&-2B h_{,\tilde t} 
+ {16 \pi GM B^4 \over 3V_3}\e^{-4B\tilde t} \nn
&& + {2\pi G \over 3}\left({1 \over l} 
 - 8\pi G B^4 b'\right)\left\{-4b'\left(
h_{,\tilde t\tilde t\tilde t\tilde t} 
+ 7B\tilde t h_{,\tilde t\tilde t\tilde t} 
+ 18B^4 h_{,\tilde t\tilde t} 
+ 24 B^3 h_{,\tilde t} 
\right)\right. \nn
&& \left. + 4(b+b')\left(h_{,\tilde t\tilde t\tilde t\tilde t} 
+ 7Bt h_{,\tilde t\tilde t\tilde t} 
+ 12 B^2h_{,\tilde t\tilde t}\right)\right\}\ .
\eea
The solution of (\ref{dS15}) is given by 
\bea
\label{dS17}
h&=&h_0\e^{-4B\tilde t} + \mbox{$M$ independent terms} \nn
h_0&\equiv& {{16 \pi GM \over 3V_3} \over
{800\pi Gb \over 3}\left({1 \over l} 
 - 8\pi G B^4 b'\right) - {8 \over B^3}}\ .
\eea
The $M$ independent terms vanish if we impose a condition 
that $h$ vanishes when $E=0$. As the solution seems to exist 
consistently, there will be (approximate) de Sitter like 
solutions  for a range of the parameters. 
Therefore the quantum correction seems to induce the de 
Sitter like brane not only in case of $M=0$ but even in 
the case of  $M\neq 0$.

For the above obtained perturbative solution (\ref{dS13}), 
Eq.(\ref{e11bb}) has the following form:
\be
\label{e11bb3}
\rho+p=-{1 \over 4\pi G_4}{1 \over h_{,\tilde t\tilde t}}\ .
\ee
If we take $M$-independent terms in (\ref{dS17}) to vanish, 
(\ref{e11bb3}) has the following form: 
\be
\label{e11bb4}
\rho+p=-{1 \over 4\pi G_4}{1 \over 16B^2 h_0}\ .
\ee
By using the expression of $h_0$ in (\ref{dS17}), the weak energy 
condition is satisfied, at least for large $\tilde t$, if
\be
\label{e11bb5}
{800\pi Gb \over 3}\left({1 \over l} 
 - 8\pi G B^4 b'\right) - {8 \over B^3}<0\ .
\ee
This depends on the field contents on the brane. 

\section{Properties of inflationary brane in AdS-Schwarzschild 
Black Hole}

Let us briefly discuss the properties of the found inflationary brane 
universe. When $a$ is large, the metric 
(\ref{AdSS}) has the following form:
\be
\label{eq13} 
ds_{\rm AdS-S}^2 \rightarrow {a^2 \over l^2}\left(dt^2 
+ l^2 d\Omega_3^2\right)\ ,
\ee
which tells that the CFT time $t_{\rm CFT}$ is equal 
to the AdS time $t$ times the factor ${a \over l}$, 
at least when the radius of the brane is large enough:
\be
\label{eq14}
t_{\rm CFT}={a \over l}t\ .
\ee
Therefore the energy $E_{\rm CFT}$ in CFT is related 
with the energy $E_{\rm AdS}$ in AdS by \cite{SV}
\be
\label{eq15}
E_{\rm CFT}={l \over a}E_{\rm AdS}\ .
\ee
The factor ${l \over a}$ in front of Eqs.(\ref{e10b}) and 
(\ref{e11b}) appears due to the above scaling of the energy 
in (\ref{eq15}) or time in (\ref{eq14}). 

The AdS$_5$-Schwarzschild black hole solution in (\ref{AdSS}) 
has a horizon at $a=a_H$, where $h(a)$ vanishes \cite{HP}:
\be
\label{eq16}
h(a_H)= {a_H^2 \over l^2} + 1 - {16\pi GM \over 3 V_3 a_H^2}=0\ .
\ee
Then considering the moment the brane crosses these points and
 using (\ref{e7}),  one gets
\bea
\label{e17}
&& H = \pm \left[{1 \over l} + {\pi G \over 3}\left\{ 
-4b'\left(\left(H_{,\tilde t \tilde t \tilde t} + 4 H_{,\tilde t}^2 
+ 7 H H_{,\tilde t\tilde t} + 18 H^2 H_{,\tilde t} + 6 H^4\right) 
\right.\right.\right. \nn
&&\ \left. + {4 \over a_H^2} \left(H_{,\tilde t} + H^2\right)\right) 
+ 4(b+b') \left(\left(H_{,\tilde t \tilde t \tilde t} 
+ 4 H_{,\tilde t}^2 \right.\right. \nn
&&\ \left. + 7 H H_{,\tilde t\tilde t} 
+ 12 H^2 H_{,\tilde t} \right) 
\left.\left.\left. - {2 \over a_H^2} \left(H_{,\tilde t} 
+ H^2\right)\right) \right\}\right]\ .
\eea
The sign $\pm$ depends on whether the brane is expanding or 
contracting. 
Obviously, if the higher derivative of the Hubble constant $H$ is 
large, the quantum correction becomes essential. 

We now assume that the brane behaves as de Sitter (inflationary) 
space in (\ref{dS1}) near the horizon. As it has been shown, 
this is quantum-induced brane Universe. (Parameters 
$B$, $A$ are defined by quantum effects).
Note that this is not the
exact solution for positive 
(non-vanishing) black hole mass $M>0$  while there can be 
another kind of solution, not of de Sitter type.
The above assumption, however, is not so unnatural since 
it only means that the brane universe expands (or shrinks) 
uniformly near the horizon.
Then Eqs.(\ref{eq16}) and (\ref{e17}) 
have the following forms: 
\bea
\label{eq16b}
h(a_H) &=& {A^2\cosh^2 \tilde t_H \over l^2} + 1
 - {16\pi GM \over 3 V_3 A^2\cosh^2 \tilde t_H }=0 \\
\label{e17b}
H &=& \pm \left[{1 \over l} + {\pi G \over 3}\left\{ 
-4b'\left(-4\left(B^4 - {B^2 \over A^2}\right)
{1 \over \cosh^2 \tilde t_H} + 6B^4\right) \right.\right. \nn
&& \left. + 8(b + b') \left(B^4 - {B^2 \over A^2}\right)
{1 \over \cosh^2 \tilde t_H}\right]\ .
\eea
Here the brane crosses the horizon when $\tilde t=\tilde t_H$. 
Thus, quantum-corrected Hubble parameter at the horizon is defined.
The quantum correction becomes large when the rate $B$ of expansion 
of the universe is large. 

Let the entropy ${\cal S}$ of CFT on the brane is given 
by the Bekenstein-Hawking entropy of the AdS$_5$ black hole 
${\cal S}={V_H \over 4G}$. 
Here $V_H$ is the area of the horizon, which is equal to the 
spatial brane when the brane crosses the horizon:
$V_H=a_H^3 V_3$. 
If the total entropy ${\cal S}$ is constant during the 
cosmological evolution, the entropy density $s$ is given by (see \cite{SV})
\be
\label{e20}
s={{\cal S} \over a^3 V_3}= {l a_H^3 \over 2G_4 a^3}\ .
\ee
Here  Eq.(\ref{e12}) is used. The expression in (\ref{e20}) 
is identical with the classical one. The quantum correction 
appears when we express $s$ in terms of the quantities in 
brane universe, say $H$, $H_{,\tilde t}$ etc., by using 
(\ref{e17}). 

The Hawking temperature of the black hole is given by (see \cite{SV}) 
\be
\label{e21}
T_H={h'(a_H) \over 4\pi} 
={a_H \over 2\pi l^2} + {8GM \over 3V_3 a_H^3} 
={a_H \over \pi l^2} + {1 \over 2\pi a_H}\ .
\ee
Here (\ref{eq16}) is used. 
As in (\ref{eq14}), the temperature $T$ on the brane 
is different from that of AdS$_5$ by the factor 
${l \over a}$:
\be
\label{e22}
T={l \over a}T_H
={a_H \over \pi a l} + {l \over 2\pi a a_H}\ .
\ee
Especially when $a=a_H$
\be
\label{e23}
T={1 \over \pi l} + {l \over 2\pi a_H^2}\ .
\ee
Then from (\ref{e9}) and (\ref{e17}), one gets
\bea
\label{e9b}
T  &=&  {l \over \pi}\left[ - 2H_{,\tilde t} 
\pm {\pi G \over 3 }\left\{ 
-4b'\left(\left(H_{,\tilde t \tilde t \tilde t \tilde t} 
+ 15 H_{,\tilde t} H_{\tilde t\tilde t} 
+ 7 H H_{,\tilde t\tilde t\tilde t}  
\right. \right.\right.\right. \nn
&& \left.\left. + 18 H^2 H_{,\tilde t\tilde t}
+ 36 H H_{,\tilde t}^2 + 24 H^3 H_{,\tilde t} \right) + {4 \over a^2} 
\left(H_{,\tilde t\tilde t}  - 2 H^3\right) \right) \nn
&& + 4(b+b') \left(\left(H_{,\tilde t \tilde t \tilde t \tilde t} 
+ 15 H_{,\tilde t} H_{,\tilde t\tilde t} 
+ 7 H H_{,\tilde t\tilde t\tilde t} + 12 H^2 H_{,\tilde t\tilde t}
\right.\right. \nn
&& \left.\left.\left.\left. + 24 H H_{,\tilde t}^2 \right) 
 - {2 \over a_H^2} \left(H_{,\tilde t\tilde t} 
 - 2H^2\right)\right) \right\}\right]\ .
\eea
Again, if the higher derivative of the Hubble constant $H$ is 
large, the quantum correction becomes important. 
For the solution (\ref{dS1}), we have 
\be
\label{e9bb}
T = {l\over \pi}\left[ -
2H_{,\tilde t}  \pm {\pi G \over 3 }(48b' + 16b) 
\left(B^4 - {B^2 \over A^2}\right) {\sinh B\tilde t_H 
\over \cosh^3 B\tilde t_H}\right]\ .
\ee
The quantum correction becomes dominant when $B\tilde t_H$ 
is of order 
unity but $B$ ($\neq {1 \over A}$) is large or $A$ is small. 
Since the radius of the horizon is given by 
$a_H=A\cosh B\tilde t_H$, this might mean that if quantum 
correction is large then the radius of the black hole 
is small.
Eq.(\ref{e22}) tells that the temperature on the brane 
should be determined by the Hawking temperature of the black hole. 
If there is no quantum correction, the temperature is directly 
related with $H_{,\tilde t}$ but Eq.(\ref{e9b}) or (\ref{e9bb}) 
tells that the quantum correction breaks this simple relation. 
Thus, the main qualitative role of quantum effects was to provide
the explicit inflationary brane universe solution which does not exist 
otherwise (at least, in our approach to brane-worlds).
The parameters of quantum CFT enter to the Hubble constant, Hawking 
temperature, energy (entropy) and they may change their classical values.

\section{Discussion}

In summary, it is shown that in d5 AdS black hole background, 
the inflationary 
brane induced by CFT quantum effects 
 may occur in the same way as in refs.\cite{HHR, NObr}.
It is important to note that brane stress tensor is completely 
defined by dual quantum CFT (and also probably, by brane QG) 
and it is not chosen by hands as it happens often 
in the traditional brane-world scenarios, where the brane 
tension is fine-tuned. We also investigated the energy 
conditions and found that the energy density is always 
positive but the weak energy condition might be broken by 
the quantum effects. 
When the quantum-induced brane crosses horizon of AdS BH, 
the Hubble constant, brane entropy and the Hawking temperature 
(also at the horizon) are found with account of quantum corrections.
The similarity between CFT entropy at the horizon and FRW equations 
discovered in refs.\cite{EV,SV} is extended for the presence 
of quantum effects. These results may be important for the 
generalization of cosmological entropy bounds in the case of 
quantum gravity. From another side, it would be interesting 
to use such study with the purpose of extension of AdS/CFT 
correspondence for cosmological (AdS) 
backgrounds \cite{NOA}. Clearly, there are many questions about
physical interpretation of some of the obtained results which 
should be carefully investigated in the future.

\ 

\noindent
{\bf Acknoweledgments} The work by SDO has been supported 
in part by CONACyT (CP, Ref.990356 and grant 28454E).


\begin{thebibliography}{99}
\bibitem{H} S.W. Hawking, {\sl Comm.Math.Phys.} {\bf 43} (1974) 199.
\bibitem{HS} G.'t Hooft, gr-qc/9312026;
L. Susskind, {\sl J.Math.Phys.} {\bf 36} (1995) 6337;
W. Fischler and L. Susskind, hep-th/9806039;
R. Easther and D. Lowe, hep-th/9902088, {\sl Phys.Rev.Lett.} 
{\bf 82} (1999) 4967;
D. Bak and S. Rey, hep-th/9902173, {\sl Class.Quant.Grav.} {\bf 17}
(2000) L83; N. Kaloper and A. Linde, hep-th/9904120;
R. Bousso, hep-th/9905177, {\sl JHEP} {\bf 9907} (1999) 004;
B. Wang, E. Abdalla and T. Osada, {\sl Phys.Rev.Lett.} {\bf 85} (2000) 5507.
\bibitem{HMS} S.W. Hawking, J. Maldacena and A. Strominger, hep-th/0002145.
\bibitem{EV} E. Verlinde, hep-th/0008140.
\bibitem{AdS}  J.M. Maldacena, 
{\sl Adv.Theor.Math.Phys.} {\bf 2} (1998) 231;
E. Witten, {\sl Adv.Theor.Math.Phys.} {\bf 2} (1998) 253;
S. Gubser, I. Klebanov and A. Polyakov, {\sl Phys.Lett.} 
{\bf B428} (1998) 105;
\bibitem{related} D. Kutasov and F. Larsen,hep-th/0009244;
F.-L. Lin, hep-th/0010127;
B. Wang, E. Abdalla and R.-K. Su, hep-th/0101073;
Y. S. Myung, hep-th/0102184;  
R.-G. Cai, hep-th/0102113;
R. Brustein, S. Foffa and G. Veneziano, hep-th/0101083;
A. Biswas and S. Mukherji, hep-th/0102138.
\bibitem{NO} S. Nojiri and S.D. Odintsov, hep-th/0011115.
\bibitem{Gubser} S.S. Gubser, {\sl Phys.Rev.} {\bf D63} 
(2001) 084017, hep-th/9912001.  
\bibitem{SV} I. Savonije and E. Verlinde, hep-th/0102042.
\bibitem{RS} L. Randall and R. Sundrum, 
{\sl Phys.Rev.Lett.} {\bf 83} (1999) 3370, hep-th/9905221; 
{\sl Phys.Rev.Lett.} 
{\bf 83} (1999)4690, hep-th/9906064. 
\bibitem{HHR} S.W. Hawking, T. Hertog and H.S. Reall, 
hep-th/0003052, {\sl Phys.Rev.} {\bf D62} (2000) 043501. 
\bibitem{NObr}
S. Nojiri, S.D. Odintsov and S. Zerbini, hep-th/0001192,
{\sl Phys.Rev.} {\bf D62} (2000) 064006;
S. Nojiri and S.D. Odintsov, hep-th/0004097, 
{\sl Phys.Lett.} {\bf B484} (2000) 119.
\bibitem{extension} L. Anchordoqui, C. Nunez and K. Olsen, hep-th/0007064;
K. Koyama and J. Soda, hep-th/0101164;
T. Shiromizu and D. Ida, hep-th/0102035.
\bibitem{NOO}
S. Nojiri, O. Obregon and S.D. Odintsov, hep-th/0005127,
{\sl Phys.Rev.} {\bf D62} (2000) 104003;
S. Nojiri, O. Obregon, S.D. Odintsov and V.I. Tkach, hep-th/0101003;
S. Nojiri and S.D. Odintsov, hep-th/0102032.
\bibitem{NOZcas} S. Nojiri, S.D. Odintsov and S. Zerbini, 
hep-th/0006115, {\sl Class.Quant.Grav.} {\bf 17} (2000) 4855. 
\bibitem{HP} S. Hawking and D. Page, {\sl Comm.Math.Phys.} 
{\bf 87} (1983) 577.
\bibitem{NOA} S. Nojiri and S.D. Odintsov, hep-th/0008160, 
{\sl Phys.Lett.} {\bf B494} (2000) 135.
\bibitem{S} A. Starobinsky, {\sl Phys.Lett.} {\bf B91} (1980) 99.
\bibitem{HHR2} S.W. Hawking, T. Hertog and H.S. Reall, 
hep-th/0010252;
K. Hamada, hep-th/0101100.
\end{thebibliography}
\end{document}